# Nonsemisimple Sugawara constructions

José M. Figueroa-O'Farrill[†]

and

Sonia Stanciu[‡§]

*Department of Physics, Queen Mary and Westfield College*
*Mile End Road, London E1 4NS, UK*

## ABSTRACT

By a Sugawara construction we mean a generalized Virasoro construction in which the currents are primary fields of conformal weight one. For simple Lie algebras, this singles out the standard Sugawara construction out of all the solutions to the Virasoro master equation. Examples of nonsemisimple Sugawara constructions have appeared recently. They share the properties that the Virasoro central charge is an integer equal to the dimension of the Lie algebra and that they can be obtained by high-level contraction of reductive Sugawara constructions: they thus correspond to free bosons. Exploiting a recent structure theorem for Lie algebras with an invariant metric, we are able to unify all the known constructions under the same formalism and, at the same time, to prove several results about the Sugawara constructions. In particular, we prove that all such constructions factorize into a standard (semisimple) Sugawara construction and a nonsemisimple one (with integral central charge) of a form which generalizes the nonsemisimple examples known so far.

---

[†] e-mail: jmf@strings1.ph.qmw.ac.uk

[‡] e-mail: s.stanciu@qmw.ac.uk

[§] On leave of absence from Physikalischem Institut der Universität Bonn

## §1 Introduction

There has been a lot of interest recently on WZW models based on non-abelian nonsemisimple Lie groups [1] [2] [3] [4] [5], particularly because they allow the construction of exact string backgrounds. Nappi and Witten showed in [1] that a WZW model based on a central extension of the two-dimensional euclidean group describes the homogeneous four-dimensional spacetime corresponding to a gravitational plane wave. For a standard WZW model, the nonperturbative proof of its conformal invariance relies on the Sugawara construction. But since the group considered in [1] is not semisimple, the Killing metric of its Lie algebra is degenerate and the standard Sugawara construction does not exist. Nappi and Witten realized that there is another invariant bilinear form on the Lie algebra which is nondegenerate and which allows for a Sugawara-type construction. In other words, using this other bilinear form, one constructs a solution of the Virasoro master equation [6] with the added property that all the currents are primary fields of conformal weight one.

The construction in [1] was quickly generalized by Sfetsos [3] to the abelian extensions of the $d$-dimensional euclidean algebra. In this case the resulting conformal field theory has central charge $c = d^2$–again equal to the dimension of the algebra. A larger class of examples of nonsemisimple Sugawara constructions was found by Olive, Rabinovici, and Schwimmer [4] by a high-level contraction of ordinary Sugawara constructions associated to reductive Lie algebras. As a special result of their construction they recovered the example in [1] and we will see that the examples in [3] also follow from this construction. Since for high level the affine algebra becomes simply a direct product of affine $u(1)$'s, the central charge of the Virasoro generator is integral in all their examples and, moreover, the energy momentum tensor can be seen to be that of free scalar fields. For the Nappi-Witten example, this fact corroborates the observation that the string background described in [1] is connected by a duality transformation to four-dimensional Minkowski spacetime [7] [2] [3].

In [5] Mohammedi spelled out the conditions for a Lie algebra to admit a Sugawara construction. He found that asking for a Sugawara construction is equivalent to demanding that the Lie algebra possess an invariant metric. He also obtained a formula for the central charge of the Sugawara construction from where one sees that the integrality of the central charge is not an a priori consequence of the construction. He thus emphasized the possibility of obtaining a non-integral Virasoro central charge from a nonsemisimple Sugawara construction. The search for new nonsemisimple examples with a non-integral central charge was the main motivation for the present paper. We will prove however that they do not exist. In other words, non-integral values of the central charge can always be traced back to a standard Sugawara construction for a semisimple factor.

What makes our results possible is a structure theorem due to Medina and Revoy [8] for Lie algebras admitting an invariant metric. The punch line is that the class of such algebras is the smallest class containing the simple Lie algebras and abelian Lie algebras and which is closed under the operations of taking direct sum and "double extension" (see below for a precise definition). In other words, the theorem states that one can obtain all Lie algebras with an invariant metric starting from the simple Lie algebras and the one-dimensional algebra by iterating those two operations. In this paper we shall explore the consequences of this theorem for the nonsemisimple Sugawara constructions and in the process we will be able to refine the structure theorem considerably. In particular, we shall be able to obtain some control on the Virasoro central charge of the construction. Our main result is that the general Sugawara construction will factorize into a standard semisimple Sugawara construction and one with integral central charge. We will limit ourselves to a summary of results—leaving a fuller, more detailed treatment for a forthcoming publication.

This note is organized as follows. After reviewing the result of [5] concerning the existence of the Sugawara construction, we discuss briefly the notion of a double extension [8]. We then proceed to show how the known examples can be understood as double extensions. They will all turn out to be double extensions of abelian Lie algebras. We then turn to some more general considerations and investigate how the Virasoro central charge behaves under the process of double extension. We will find that when the algebra we double extend is solvable (for example, abelian), then the central charge will be integral. This forces us to consider the double extension of Lie algebras which have a semisimple factor. We show however that the semisimple factor survives the double extension unperturbed, and from this conclude that the most general Sugawara construction is a sum of a semisimple Sugawara construction with a nonsemisimple factors each of which with integral central charge.

§2 THE GENERAL SUGAWARA CONSTRUCTION

Let us start by reviewing the general Sugawara construction in [5]. Let $\mathfrak{g}$ be a finite-dimensional real Lie algebra. By a Sugawara construction for $\mathfrak{g}$ we mean the construction of a Virasoro algebra out of (normal-ordered) bilinears in the currents of $\mathfrak{g}$, with the extra property that the currents are primary fields of weight one. In fact this latter condition is sufficient, as we shall see presently. For $\mathfrak{g}$ a simple algebra, this condition uniquely singles out the standard Sugawara construction. By adding the Sugawara construction for each simple factor, one can extend this to semisimple Lie algebras; while abelian algebras are taken care of by their equivalence to free bosons. When a Lie algebra is not a direct sum of algebras of these types, the Sugawara construction is not guaranteed to exist. As was shown in [5] and as we shall now see, the necessary and sufficient condition for a Lie algebra to admit a Sugawara construction is that it admit an invariant nondegenerate symmetric bilinear form—that is, an invariant metric—, and provided that the central extension of the current algebra is fixed appropriately.

Choose a basis $\{X_i\}$ for $\mathfrak{g}$ and fix the following current algebra

$$X_i(z)X_j(w) = \frac{g_{ij}}{(z-w)^2} + \frac{f_{ij}{}^k X_k(w)}{z-w} + \text{reg} , \qquad (2.1)$$

where $f_{ij}{}^k$ are the structure constants of $\mathfrak{g}$. Associativity of the above OPE forces $g_{ij}$ to be an invariant bilinear form. Let $T(z) \equiv \Omega^{ij}(X_iX_j)(z)$, for some symmetric bivector $\Omega^{ij}$ and where the parentheses indicates normal ordering according to the standard point-splitting procedure. We now investigate the conditions on our parameters $g_{ij}$ and $\Omega^{ij}$, such that $T$ obeys the Virasoro algebra and such that every current $X_i(z)$ is a primary field of conformal weight one. This latter condition translates into

$$X_i(z)T(w) = \frac{X_i(w)}{(z-w)^2} + \text{reg} , \qquad (2.2)$$

and a short calculation now shows that if $T$ obeys (2.2) then it also obeys the Virasoro algebra with central charge $c = 2\Omega^{ij}g_{ij}$. The condition (2.2) translates into the following conditions. First of all, the vanishing of the first-order pole says that $T$ commutes with the charges associated to the currents $X_i$ whence the bivector $\Omega^{ij}$ must be invariant: $f_{lj}{}^m \Omega^{ij} = -f_{lj}{}^i \Omega^{jm}$. Furthermore, the second-order pole equation becomes

$$2\Omega^{mj}g_{kj} + \Omega^{ij}f_{ki}{}^l f_{lj}{}^m = \delta^m_k , \qquad (2.3)$$

which using the invariance of $\Omega^{ij}$, can be rewritten as

$$\Omega^{mj}(2g_{kj} + \kappa_{kj}) = \delta^m_k , \qquad (2.4)$$

where $\kappa_{ij} \equiv f_{ik}{}^l f_{jl}{}^k$ is the Killing form. In other words, $\Omega^{ij}$ is invertible with inverse $\Omega_{ij} = \kappa_{ij} + 2g_{ij}$.

In summary, a Lie algebra $\mathfrak{g}$ admits a Sugawara construction provided that it possesses an invariant metric $\Omega_{ij}$ and provided that the currents obey the algebra (2.1) with $g_{ij} = \frac{1}{2}(\Omega_{ij} - \kappa_{ij})$. In addition, the Virasoro central charge is given by

$$c = \dim \mathfrak{g} - \Omega^{ij}\kappa_{ij} . \qquad (2.5)$$

In the special case of $\mathfrak{g}$ abelian, the Killing form is identically zero and we recover the free boson construction with $c = \dim \mathfrak{g}$. Similarly if $\mathfrak{g}$ is simple, $\Omega^{ij}$ is a multiple of the Killing form, say $\Omega_{ij} = \mu \kappa_{ij}$. The usual formulas are then recovered by taking $\mu = (x + 2g^*)/2g^*$ with $g^*$ the dual Coxeter number and $x$ the level.

With this result in mind, it thus behooves us to investigate under which conditions a Lie algebra admits an invariant metric. The problem of determining the structure of Lie algebras with invariant metrics has been studied by Medina and Revoy [8], who proved a structure theorem for these algebras. It is clear that if $\mathfrak{g}_1$ and $\mathfrak{g}_2$ are Lie algebras with invariant metrics so is their direct product $\mathfrak{g}_1 \times \mathfrak{g}_2$ with the direct product metric. A Lie algebra with invariant metric is called *indecomposable* if it cannot be written as such a direct product. It is clear that simple Lie algebras and the 1-dimensional Lie algebra are indecomposable; but as the example of [1] shows, these are not the only ones. To obtain further examples of indecomposable algebras it turns out to be sufficient to consider an operation known as a "double extension." This is a generalization of the following more familiar construction.

Let $\mathfrak{h}$ be any Lie algebra and let $\mathfrak{h}^*$ denote its dual. A basis $\{H_a\}$ for $\mathfrak{h}$ induces a canonical dual basis $\{H^a\}$ obeying $\langle H^a, H_b \rangle = \delta^a_b$. Since $\mathfrak{h}$ acts on $\mathfrak{h}^*$ via the coadjoint representation, we can define on the vector space $\mathfrak{h} \oplus \mathfrak{h}^*$ the structure of a Lie algebra as follows. On $\mathfrak{h}$ we have the original Lie bracket $[H_a, H_b] = f_{ab}{}^c H_c$, whereas we make $\mathfrak{h}^*$ abelian: $[H^a, H^b] = 0$. The mixed brackets are given by $[H_a, H^b] = -f_{ac}{}^b H^c$—that is, the coadjoint action. The Jacobi identities are easy to check and all boil down to the Jacobi identity for $\mathfrak{h}$. The resulting algebra is the semidirect product of $\mathfrak{h}$ and $\mathfrak{h}^*$ and is written $\mathfrak{h} \ltimes \mathfrak{h}^*$. Because of the very definition of the coadjoint representation, the dual pairing $\langle,\rangle$ of $\mathfrak{h}$ and $\mathfrak{h}^*$ provides $\mathfrak{h} \ltimes \mathfrak{h}^*$ with an invariant metric. Parenthetically, the triple $(\mathfrak{h} \ltimes \mathfrak{h}^*, \mathfrak{h}, \mathfrak{h}^*)$ is a Manin triple [9] associated to the trivial bialgebra structure on $\mathfrak{h}$.

We now introduce a Lie algebra $\mathfrak{g}$ with an invariant metric. We let the invariant metric have components $\Omega_{ij}$ relative to a fixed basis $\{X_i\}$ for $\mathfrak{g}$. And we suppose that $\mathfrak{h}$ acts on $\mathfrak{g}$ in such a way that it preserves both the bracket and the metric; in other words, $\mathfrak{h}$ acts on $\mathfrak{g}$ via antisymmetric derivations. Explicitly this means that we have an action $H_a \cdot X_i = f_{ai}{}^j X_j$. The antisymmetry condition becomes

$$f_{ai}{}^k \Omega_{kj} = -f_{aj}{}^k \Omega_{ik} \tag{3.1}$$

and the derivation condition can be read from $H_a \cdot [X_i, X_j] = [H_a \cdot X_i, X_j] + [X_i, H_a \cdot X_j]$; that is,

$$f_{ij}{}^k f_{ak}{}^l = f_{ai}{}^k f_{kj}{}^l + f_{ik}{}^l f_{aj}{}^k . \tag{3.2}$$

We now define on the vector space $\mathfrak{g} \oplus \mathfrak{h} \oplus \mathfrak{h}^*$ the following Lie brackets

$$\begin{aligned}
[X_i, X_j] &= f_{ij}{}^k X_k + f_{ai}{}^k \Omega_{kj} H^a \\
[H_a, H_b] &= f_{ab}{}^c H_c \\
[H_a, X_i] &= f_{ai}{}^j X_j \\
[H_a, H^b] &= -f_{ac}{}^b H^c .
\end{aligned} \tag{3.3}$$

Antisymmetry may be in question for the $[X_i, X_j]$ bracket, but follows from (3.1), whereas the Jacobi identities for (3.3) follow trivially from the Jacobi identities of $\mathfrak{g}$ and $\mathfrak{h}$ and from (3.2). Notice that the subalgebra spanned by $\mathfrak{g} \oplus \mathfrak{h}^*$ is the abelian extension of $\mathfrak{g}$ by $\mathfrak{h}^*$, whereas the full algebra is the semidirect product of $\mathfrak{h}$ by this abelian extension.

Notice that $\mathfrak{h}^*$ is an abelian ideal of (3.3), whence the Lie algebra defined by (3.3) is not semisimple. In particular this means that its Killing form is degenerate. Nevertheless, it has the remarkable property that it does admit an invariant metric. In fact a whole family of them. To construct them, let $\Omega_{ab}$ denote any (possibly degenerate) invariant bilinear form in $\mathfrak{h}$. Relative to the given basis for $\mathfrak{g} \oplus \mathfrak{h} \oplus \mathfrak{h}^*$, let us define the following bilinear form

$$\Omega_{IJ} = \begin{array}{c} \\ X_i \\ H_a \\ H^a \end{array} \begin{pmatrix} X_j & H_b & H^b \\ \Omega_{ij} & 0 & 0 \\ 0 & \Omega_{ab} & \delta^b_a \\ 0 & \delta^a_b & 0 \end{pmatrix} . \tag{3.4}$$

This bilinear form in clearly nondegenerate with inverse

$$\Omega^{IJ} = \begin{pmatrix} \Omega^{ij} & 0 & 0 \\ 0 & 0 & \delta^a_b \\ 0 & \delta^b_a & -\Omega_{ab} \end{pmatrix} , \tag{3.5}$$

where $\Omega^{ij}$ is the inverse of $\Omega_{ij}$. The invariance of (3.4) follows immediately from the invariance of $\Omega_{ab}$ and $\Omega_{ij}$. Since $\Omega_{ab}$ was arbitrary, this gives us a family of invariant metrics parametrized by the invariant bilinear forms of $\mathfrak{h}$.

The algebra (3.3) with metric (3.4) is called the *double extension of $\mathfrak{g}$ by $\mathfrak{h}$*. For lack of a standard notation we will refer to it as $\mathfrak{h} \ltimes (\mathfrak{g} \times_c \mathfrak{h}^*)$ with the subscript on the $\times$ serves to remind us that it is not a direct product but rather a central extension. Similarly we can write the double extension in the form $(\mathfrak{h} \ltimes \mathfrak{g}) \ltimes_a \mathfrak{h}^*$, where the subscript now tells us that it is not generally a semidirect product, but rather a (not necessarily split) abelian extension. We hasten to add that these notations are not at all standard, but we find them useful mnemonic tools. Notice that if we take $\mathfrak{g} = \mathbf{0}$ then we recover the previous example $\mathfrak{h} \ltimes \mathfrak{h}^*$. Similarly, for any $\mathfrak{g}$ if the action of $\mathfrak{h}$ on $\mathfrak{g}$ is trivial, then we recover the direct product algebra $(\mathfrak{h} \ltimes \mathfrak{h}^*) \times \mathfrak{g}$.

We are now in a position to state the structure theorem of Medina and Revoy [8]. Every indecomposable Lie algebra with an invariant metric is either simple, 1-dimensional, or else a double extension of $\mathfrak{g}$ by $\mathfrak{h}$ where $\mathfrak{g}$ is a Lie algebra (not necessarily indecomposable) with an invariant metric and $\mathfrak{h}$ is either simple or 1-dimensional.

Several remarks are noteworthy. First and foremost, nothing in the results of [8] suggests that all double extensions by a simple or 1-dimensional Lie algebra yield indecomposable objects. In fact, we will see later that the double-extension of a semisimple Lie algebra is always decomposable.

A second remark, is that it should be noticed that the data one needs to construct the double extension consists of the relevant Lie algebras $\mathfrak{g}$ and $\mathfrak{h}$ and, *in addition* the invariant metric chosen for $\mathfrak{g}$. This is evident from the expression for $[X_i, X_j]$ in (3.3). In fact, a closer look reveals that it depends only on the conformal class of $\Omega_{ij}$, for if we were to rescale $\Omega_{ij}$, we can reabsorb this by inversely rescaling the generators of $\mathfrak{h}^*$. In other words, a given double extension has generically at least a two-parameter family of invariant metrics: a scale ($\Omega_{ij}$) and a translation ($\Omega_{ab}$).

## §4 Double extensions of abelian Lie algebras

Before discussing the general consequences that this structure theorem has for the Sugawara construction, let us gain some intuition by recasting the known examples in the form of double extensions. The first known examples are the ones of Nappi-Witten [1] and the generalization due to Sfetsos [3], which can both be seen as special cases of the examples obtained by Olive-Rabinovici-Schwimmer [4] via contractions. These in turn are examples of double-extensions of abelian algebras.

### The examples of Nappi-Witten and Sfetsos

Nappi and Witten considered the universal central extension of the euclidean algebra in two dimensions with generators $\{J, P_i, T\}$ and with nonzero Lie brackets $[J, P_i] = \epsilon_{ij} P_j$, $[P_i, P_j] = \epsilon_{ij} T$. Comparing with equation (3.3) we immediately recognize this Lie algebra as the double extension of the algebra of two-dimensional translations by $u(1)$. According to equation (3.4), it has a one-parameter ($b$) family of invariant metrics with the following nonzero entries: $\langle J, J \rangle = b$, $\langle P_i, P_j \rangle = \delta_{ij}$, and $\langle J, T \rangle = 1$.

From the point of view of double extensions, there is an obvious generalization. Take any real abelian Lie algebra $\mathfrak{g}$ of dimension $d$. Any metric is invariant and can be brought to a diagonal form such that all its eigenvalues are $\pm 1$. Assume that there are $p$ positive eigenvalues and $q = d - p$ negative eigenvalues. Since $\mathfrak{g}$ is abelian, the antisymmetric derivations will be $so(p, q)$.

So if $\mathfrak{h}$ is any subalgebra of $so(p,q)$ we can form the double extension of $\mathfrak{g}$ by $\mathfrak{h}$. In particular, we can take $\mathfrak{h} = so(p,q)$. The resulting algebra will be the abelian extension of the pseudo-euclidean algebra $so(p,q) \ltimes \mathfrak{g}$ by $so(p,q)^*$. In the euclidean case, taking $p = d$ and $q = 0$, we recover the examples of Sfetsos, the case $d = 2$ being the one of Nappi and Witten.

### The examples of Olive-Rabinovici-Schwimmer

The examples of Olive-Rabinovici-Schwimmer [4] can also be described in terms of double extensions of abelian algebras. Let $\mathfrak{g}$ be a semisimple Lie algebra and $\mathfrak{h}$ a subalgebra reductive in $\mathfrak{g}$ and such that there exists an $\mathfrak{h}$-invariant metric in $\mathfrak{g}$ relative to which $\mathfrak{g}$ splits as $\mathfrak{g} = \mathfrak{h} \oplus \mathfrak{h}^\perp$. This is true, for example, if $\mathfrak{g}$ is the Lie algebra of a compact Lie group and $\mathfrak{h}$ the Lie algebra of a (compact) Lie subgroup as in [4], or if both $\mathfrak{g}$ and $\mathfrak{h}$ are semisimple, among other cases. The models of [4] will be obtained as Wigner contractions of $\mathfrak{g} \times \mathfrak{h}$. Let $\mathfrak{k} = \mathfrak{h}^\perp$ denote the orthogonal complement of $\mathfrak{h}$ in $\mathfrak{g}$ relative to any $\mathfrak{h}$-invariant metric. Let us define subspaces $\mathfrak{h}_\pm \subset \mathfrak{h} \times \mathfrak{h}$ by $\mathfrak{h}_\pm = \{(h, \pm h) \in \mathfrak{h} \times \mathfrak{h}\}$. Choose bases $\{H_a^{(1)}\}$ and $\{H_a^{(2)}\}$ for the two copies of $\mathfrak{h}$ respectively, and $\{K_i\}$ for $\mathfrak{k}$. Then $\{H_a^\pm = H_a^{(1)} \pm H_a^{(2)}\}$ are bases for $\mathfrak{h}_\pm$, and the nonzero brackets of $\mathfrak{g} \times \mathfrak{h}$ are in this basis given by

$$\begin{aligned} [K_i, K_j] &= f_{ij}{}^k K_k + \tfrac{1}{2} f_{ij}{}^a \left( H_a^+ + H_a^- \right) \\ [H_a^\pm, H_b^\pm] &= f_{ab}{}^c H_c^+ \\ [H_a^\pm, K_i] &= f_{ai}{}^j K_j \\ [H_a^+, H_b^-] &= f_{ab}{}^c H_c^- \ , \end{aligned} \quad (4.1)$$

where the $f$'s are the structure constants of $\mathfrak{g}$ in the chosen basis. Comparing with (3.3) we notice some similarities, but the algebras do not quite agree. In particular, we would like that $[K_i, K_j]$ would not to close into $H_a^+$, and that $[H_a^-, H_b^-]$ and $[H_a^-, K_i]$ would vanish. The way out is to perform a contraction. To this effect, we define the following rescaled generators $H_a^\pm(\epsilon) = \epsilon^{\Delta_\pm} H_a^\pm$, and $K_i(\epsilon) = \epsilon^\Delta K_i$. Rewriting (4.1) in terms of the rescaled generators, we notice that we can get rid of the unwanted terms in the limit $\epsilon \to 0$ provided that we choose $\Delta_+ = 0$ and $\Delta_- = 2\Delta > 0$. Let us then choose these scaling dimensions, take the limit $\epsilon \to 0$ and introduce generators $X_i = K_i(0)$, $H_a = H_a^+(0)$, and $H^a = \frac{1}{2} g^{ab} H_b^-(0)$, where $g^{ab}$ is the inverse of the $\mathfrak{h}$-invariant metric on $\mathfrak{h}$. With this notation and using the invariance of the metric, we find that the algebra becomes precisely (3.3), but with $f_{ij}{}^k = 0$. In other words, the models obtained in this fashion are such that the algebra we double-extend is abelian. Notice that if we take $\mathfrak{g}$ to be any one of the de Sitter algebras for a spacetime of signature $(p, q)$—that is, $\mathfrak{g} = so(p+1, q)$ or $\mathfrak{g} = so(p, q+1)$—and $\mathfrak{h} = so(p, q)$, then we recover precisely the Nappi-Witten-Sfetsos models discussed in the previous section. For $(p, q) = (2, 0)$ this was already pointed out in [4].

In all these examples—which encompass all the cases that have been hitherto studied in the literature—the Virasoro central charge is integral and equal to the dimension of the algebra. This is easy to see from the construction in [4] because the contraction at the level of the current algebra involves taking the level of $\mathfrak{g}$ (resp. $\mathfrak{h}$) to positive (resp. negative) infinity, in which limit the current algebra becomes that of free bosons—albeit some of them timelike. In view of equation (2.5), the integrality of the Virasoro central charge is not an a priori consequence of the construction, which—since $\Omega^{ij}$ can be rescaled and still be nondegenerate—can allow for arbitrary central charge as long as the quantity $\Omega^{ij}\kappa_{ij}$ is nonzero. Indeed, as we will see shortly, from the point of view of the double extension we can understand the integrality of the Virasoro central charge in [4] as a consequence of the fact that we are double-extending an abelian algebra.

As this paper was being written, a new paper [10] appeared in which, among other things, the examples of [4] are slightly generalized. The new examples are obtained again from high-level contractions so the central charge is integral. The reason again is that one is double-extending an abelian algebra. Indeed, if $\mathfrak{g}$, $\mathfrak{h}$, and $\mathfrak{k}$ are as in the examples of [4] discussed above, we can consider the double extension of $\mathfrak{k} \times \mathfrak{h}^*$ by $\mathfrak{h}$ where $\mathfrak{h}^*$ is understood as an abelian Lie algebra. This algebra is precisely the one discussed in the appendix B of [10].

§5  THE VIRASORO CENTRAL CHARGE

We now return from the study of concrete examples to a more general study of the Virasoro central charge for the general Sugawara construction and in particular on how it behaves under double extension.

Suppose that $\mathfrak{a}$ is a Lie algebra with an invariant metric. If it is decomposable, say, $\mathfrak{a} = \bigoplus_i \mathfrak{a}_i$, the central charge is given by the sum of the central charges of each of the indecomposable factors $\mathfrak{a}_i$. Therefore without loss of generality we can restrict ourselves to $\mathfrak{a}$ indecomposable. According to the structure theorem of Medina-Revoy, we know that $\mathfrak{a}$ is either 1-dimensional, simple, or a double extension. If it is 1-dimensional, the Sugawara construction is simply that of a free boson and hence $c = 1$. If it is simple, we are in the well-trodden territory of the standard Sugawara construction. The only novelty occurs when $\mathfrak{a}$ is a double extension of $\mathfrak{g}$ by $\mathfrak{h}$. Let us compute the central charge in this case.

We use the notation of (3.3) and (3.4) to compute the coefficients $\kappa_{IJ}$ of the Killing form of the double extension $\mathfrak{h} \ltimes (\mathfrak{g} \times_c \mathfrak{h}^*)$. By definition, $\kappa_{IJ} = f_{IK}{}^L f_{JL}{}^K$. A simple calculation yields the following components

$$\begin{aligned}
\kappa_{ij} &= f_{ik}{}^l f_{jl}{}^k = \kappa_{ij}^{\mathfrak{g}} \\
\kappa_{ia} &= f_{ij}{}^k f_{ak}{}^j \\
\kappa_{ab} &= f_{ai}{}^j f_{bj}{}^i + 2 f_{ac}{}^d f_{bd}{}^c \\
\kappa_i{}^a &= \kappa_a{}^b = \kappa^{ab} = 0 \ .
\end{aligned} \quad (5.1)$$

The computation of the central charge now yields

$$\begin{aligned}
c(\mathfrak{a}) &= \dim \mathfrak{a} - \Omega^{IJ}\kappa_{IJ} = \dim \mathfrak{g} + 2\dim \mathfrak{h} - \Omega^{ij}\kappa_{ij}^{\mathfrak{g}} - 2\kappa_a{}^a + \Omega_{ab}\kappa^{ab} \\
&= c(\mathfrak{g}) + 2\dim \mathfrak{h} \ .
\end{aligned} \quad (5.2)$$

Therefore we see that the integrality properties of the Virasoro central charge of a double extension are unaffected by $\mathfrak{h} \oplus \mathfrak{h}^*$ which only contributes its dimension. As an example, if $\mathfrak{g}$ is solvable (in particular, abelian) $\kappa_{ij}^{\mathfrak{g}} \equiv 0$ and $c(\mathfrak{a}) = \dim \mathfrak{a}$ which explains the impossibility of obtaining non-integral central charges by the constructions in [4].

According to the structure theorem, $\mathfrak{g}$ above is any Lie algebra with an invariant metric. We can apply the structure theorem again to decompose $\mathfrak{g}$. It is going to be the orthogonal direct sum of Lie algebras which are again either 1-dimensional, simple, or double extensions. We can then apply the same procedure to decompose those which are double extensions further. This process will eventually finish, since $\mathfrak{a}$ was assumed to be finite-dimensional. It is easy to see that the contribution to the $\Omega^{IJ}\kappa_{IJ}$ term in the formula (2.5) for the Virasoro central charge will come only from the simple Lie algebras which either appear as factors or which we double-extend. In other words, it is unavoidable, if we want to obtain a non-integral central charge from a nonsemisimple Sugawara construction, that we consider a nonsemisimple algebra obtained by double-extending a semisimple algebra in some way. It is moreover clear that it is sufficient to double-extend algebras with semisimple factors, for if the algebra is "twisted" in any way, then by the structure theorem it is itself a double extension of some other algebra, and—using (5.2)—the integrality properties of the central charge are given in terms of the central charge for that other algebra. By the same token, the nonsemisimple factor need only be taken abelian, because something else would itself be a double extension and the whole thing could be understood as the double extension of an algebra with a semisimple factor. Eventually we can undo all the double extensions and be left with an algebra which is the product of a semisimple algebra and an abelian algebra; in other words, a reductive Lie algebra. We tackle this problem next.

## §6 Double extensions of reductive Lie algebras

The double extension of a reductive Lie algebra is somewhat different than extending, say, an abelian algebra. Whereas an abelian algebra $\mathfrak{g}$ has many antisymmetric derivations—in fact, the antisymmetric derivations are precisely $so(\mathfrak{g})$—a semisimple algebra has in general fewer—indeed, they are precisely the inner derivations, hence $\mathfrak{g}$. We saw when we discussed the examples of Nappi-Witten-Sfetsos that these are the maximally double-extended abelian algebras, where we double-extend by the Lie algebra of all antisymmetric derivations. Similar but smaller examples can be constructed by choosing arbitrary subalgebras of $so(\mathfrak{g})$. In this section we will analyze the double extensions of reductive Lie algebras $\mathfrak{g} = \mathfrak{s} \times \mathfrak{z}$, with $\mathfrak{s}$ semisimple and $\mathfrak{z}$ abelian. The main result is that the double extension of $\mathfrak{g}$ by any $\mathfrak{h}$ will be decomposable with $\mathfrak{s}$ as a factor.

Let $\mathfrak{g} = \mathfrak{s} \times \mathfrak{z}$ be a reductive Lie algebra. Since $[\mathfrak{s}, \mathfrak{s}] = \mathfrak{s}$, any invariant metric must be orthogonal relative to the above split: $\langle \mathfrak{s}, \mathfrak{z} \rangle = \langle [\mathfrak{s}, \mathfrak{s}], \mathfrak{z} \rangle = \langle \mathfrak{s}, [\mathfrak{s}, \mathfrak{z}] \rangle = 0$. We can choose a basis in $\mathfrak{z}$ to bring its restriction to $\mathfrak{z}$ to standard pseudo-euclidean form: diagonal with entries $\pm 1$; and the restriction of the invariant metric to each simple ideal of $\mathfrak{s}$ is a nonzero multiple of the Killing form for that ideal. Choose one such invariant metric on $\mathfrak{g}$. In order to double extend $\mathfrak{g}$ by $\mathfrak{h}$ (simple or 1-dimensional) we need to look at the antisymmetric derivations of $\mathfrak{g}$. It is easy to show that the antisymmetric derivations are precisely $\mathfrak{s} \times so(\mathfrak{z})$. Thus we need a map $\mathfrak{h} \to \mathfrak{s} \times so(\mathfrak{z})$.

Choose bases $\{X_i\}$ for $\mathfrak{s}$ and $\{Z_\alpha\}$ for $\mathfrak{z}$. Relative to these bases the metric on on $\mathfrak{s}$ is given by $\langle X_i, X_j \rangle = \Omega_{ij}$ and on $\mathfrak{z}$ is given by $\langle Z_\alpha, Z_\beta \rangle = \Omega_{\alpha\beta}$. Let $\mathfrak{h}$ have basis $\{H_a\}$ and $\mathfrak{h}^*$ have canonical dual basis $\{H^a\}$, and let the map $\mathfrak{h} \to \mathfrak{s} \times so(\mathfrak{z})$ be given by $H_a \mapsto h_a{}^i X_i + h_a{}^{\alpha\beta} M_{\alpha\beta}$, where $\{M_{\alpha\beta} = -M_{\beta\alpha}\}$ denote the generators of $so(\mathfrak{z})$. The double extension is then defined by the following brackets:

$$\begin{aligned}
[X_i, X_j] &= f_{ij}{}^k X_k + h_a{}^k f_{ki}{}^l \Omega_{lj} H^a \\
[Z_\alpha, Z_\beta] &= -2 h_a{}^{\gamma\delta} \Omega_{\alpha\gamma} \Omega_{\beta\delta} H^a \\
[H_a, X_i] &= h_a{}^j f_{ij}{}^k X_k \\
[H_a, Z_\alpha] &= 2 h_a{}^{\beta\gamma} \Omega_{\alpha\gamma} Z_\beta \\
[H_a, H_b] &= f_{ab}{}^c H_c \\
[H_a, H^b] &= -f_{ac}{}^b H^c \ .
\end{aligned} \quad (6.1)$$

Thus, if we change basis $X_i \rightsquigarrow X'_i = X_i + \Omega_{ij} h_a{}^j H^a$, we can get rid of the central extension in $\mathfrak{s}$: $[X'_i, X'_j] = f_{ij}{}^k X'_k$ at the price that $\mathfrak{s}$ is no longer orthogonal to $\mathfrak{h}$. Indeed, $\langle H_a, X'_j \rangle = \Omega_{jk} h_a{}^k$. We can regain orthogonality by redefining the generators $H_a \rightsquigarrow H'_a - h_a{}^i X'_i$. This has the added bonus that the algebra factorizes: $[H'_a, X'_i] = 0$. The only price now is that the invariant metric receives a minor modification; although it remains orthogonal:

$$\Omega_{IJ} = \begin{array}{c} \\ X'_i \\ Z_\alpha \\ H'_a \\ H^a \end{array} \begin{pmatrix} X'_j & Z_\beta & H'_b & H^b \\ \Omega_{ij} & 0 & 0 & 0 \\ 0 & \Omega_{\alpha\beta} & 0 & 0 \\ 0 & 0 & \Omega_{ab} - h_a{}^i h_b{}^j \Omega_{ij} & \delta_a^b \\ 0 & 0 & \delta_b^a & 0 \end{pmatrix} . \quad (6.2)$$

What we are left with then is the orthogonal direct sum of $\mathfrak{s}$ with the double extension of $\mathfrak{z}$ by $\mathfrak{h}$, which since $\mathfrak{z}$ is abelian has been treated already above.

Furthermore, the affine algebra (2.1) also decouples, since the $g_{IJ}$ will also be block diagonal. In the special case of $\mathfrak{s}$ simple with invariant metric $\Omega_{ij} = \mu \kappa_{ij}$ with $\kappa$ the Killing metric and $\mu$ any nonzero real number, we can recognize in the subalgebra of (2.1) generated by the currents $X'_i(z)$, the affine simple algebra $\widehat{\mathfrak{g}}$ at level $x = 2g^*(\mu - 1)$. The decomposability of the affine algebra also means that the Sugawara tensor decomposes into two commuting pieces: the standard Sugawara construction for $\widehat{\mathfrak{g}}$ at level $x$ and the contribution from the double extension of $\mathfrak{z}$ by $\mathfrak{h}$. The Virasoro central charge is easy to compute from (2.5). Again if $\mathfrak{s}$ is simple, the Virasoro central charge is given in terms of the level $x = 2g^*(\mu - 1)$ by

$$c = 2 \dim \mathfrak{h} + \dim \mathfrak{z} + \frac{x \dim \mathfrak{g}}{x + 2g^*} . \quad (6.3)$$

In other words, what we would obtain by taking the standard Sugawara construction of $\widehat{\mathfrak{g}}$ at level $x$, and a model of the kind discussed in [1], [3] and [4]. In a sense then, the Sugawara construction associated to the double-extension of a reductive Lie algebra yields nothing new. Notice that the results in this section do not depend in any essential way on $\mathfrak{z}$ being abelian. Had we taken any other algebra, we would have been able to again factor $\mathfrak{s}$ out of the double extension.

## §7 Conclusion

The above results imply a strengthening of the structure theorem of [8]. In fact, the class of (finite-dimensional, real) Lie algebras with an invariant metric is now seen to be the following: it is the product of the class of semisimple Lie algebras with the class obtained from the 1-dimensional Lie algebra under the operations of taking direct sum and double extension by simple algebras or by the 1-dimensional Lie algebra. In particular, the solvable Lie algebras are a subclass obtained by always double extending by the 1-dimensional Lie algebra. This has repercussions for the Sugawara construction. The general Sugawara construction decomposes into the sum of two commuting terms: a semisimple Sugawara construction and the Sugawara construction for nonsemisimple Lie algebras. Therefore we can recognize two kinds of "fundamental" Sugawara constructions: the one associated to simple Lie algebras and the one associated with indecomposable nonsemisimple Lie algebras with an invariant metric. Moreover, these latter construction always yields an integral central charge.

For an indecomposable nonsemisimple Lie algebra, we can define a notion of depth as follows. We "unravel" the algebra by undoing all the double extensions that we find. The number of double extensions is then the depth. The only indecomposable nonsemisimple Lie algebra with depth zero is the 1-dimensional Lie algebra; whereas the examples of [1], [3], and [4] have all depth one. The structure of "deeper" algebras may be an interesting problem to explore.

## ACKNOWLEDGEMENTS

It is a pleasure to thank Noureddine Mohammedi for the original motivation to look into the problem, and Eduardo Ramos for useful conversations. We are also grateful to Takashi Kimura and Sasha Voronov for helpful e-correspondence.